\documentstyle[11pt,newpasp,twoside,epsf]{article}
\def\edcomment#1{\iffalse\marginpar{\raggedright\sl#1\/}\else\relax\fi}
\reversemarginpar

\begin{document}
\title{Launching of resistive magnetic protostellar jets}
 \author{Miljenko \v{C}emelji\'{c}}
\affil{$^1$Astrophysikalisches Institut Potsdam, An der Sternwarte 16, 14482 Potsdam, Germany}
\author{Christian Fendt}
\affil{$^{1,2}$Universit\"at Potsdam, Institut f\"ur Physik, Am Neuen Palais 10, 14469 Potsdam, Germany}

\begin{abstract}
Turbulent accretion disks are the most probable origin of jets from young stellar objects. Using an extended version of ZEUS-3D code in the axisymmetry option we investigate the evolution of a disk wind into a collimated jet. Resistive jets propagate slower with increasing magnetic diffusivity, and the degree of collimation decreases. The resistive disk included in the simulations ejects persistent, well collimated jet.
\end{abstract}

\section{Magnetic jets}

Astrophysical jets are observed in objects with an accretion disk. The similarity of jets from the different sources implies that the basic mechanism for the jet formation should be the same.

The mechanism which turns the inflowing, accreting matter into an disk outflow  is still not understood.  However, we know that for lifting the matter into the corona magnetic fields play a major role. 
In order to model the time-dependent evolution of jet formation we solve resistive MHD equations:
\[
{\frac{\partial\rho}{\partial t}}\!+\!\nabla\cdot(\rho\vec{u})\!=\!0,\!\ e\!=\!p/(\gamma-1),\!\ \rho\left[{\frac{\partial\vec{u}}{\partial t}}\!+\!\left(\vec{u}\cdot\nabla\right)\vec{u} \right]\!+\!\nabla p\!+\!\rho\nabla\Phi\!-\!\frac{\vec{j}\times\vec{B}}{c}\!=\!0
\]
\[
{\frac{\partial\vec{B}}{\partial t}}\!-\!\nabla\times\left(\vec{u}\times\vec{B}\!-\!{\frac{4\pi}{c}}\eta\vec{j}\right)\!=\!0,\!\ \nabla\cdot\vec{B}\!=\!0,\!\ \frac{4\pi}{c}\vec{j}\!=\!\nabla \times \vec{B}
\]

\begin{figure}
\plotone{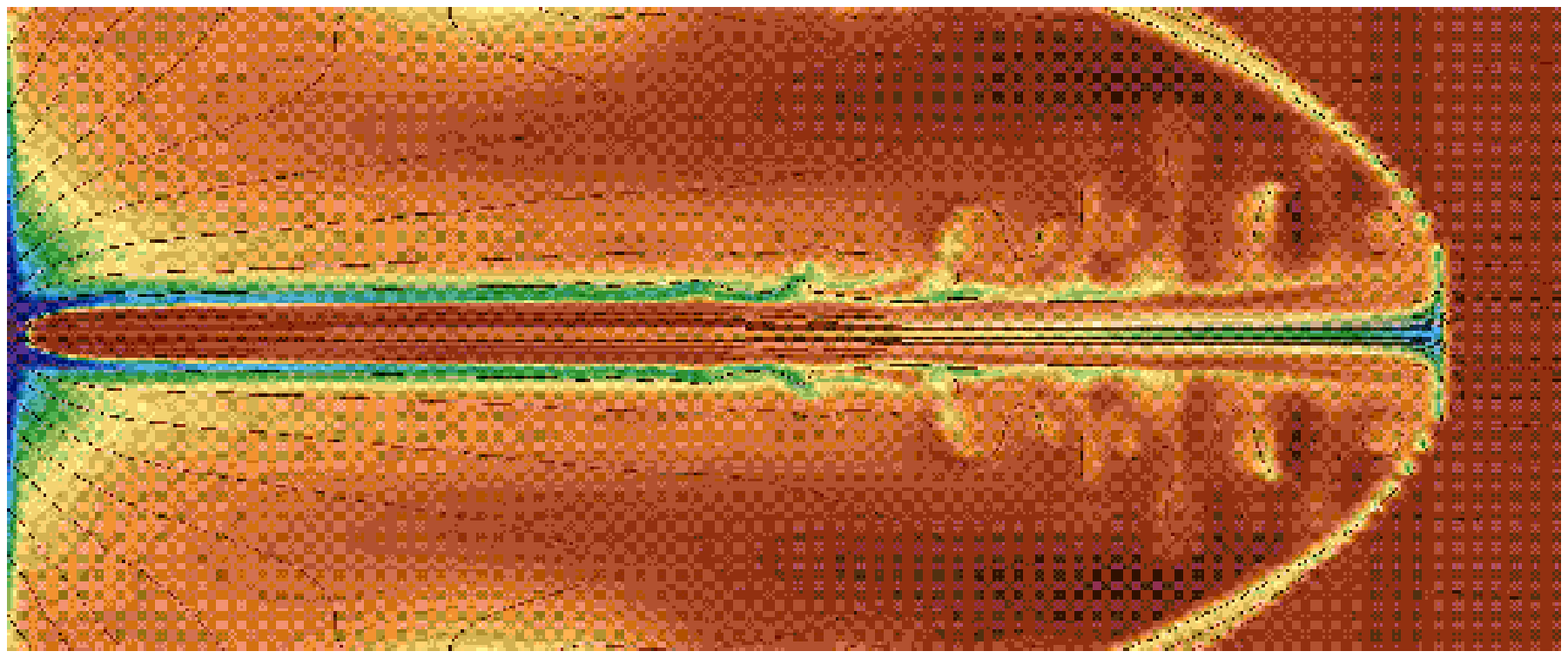}\\
\vspace{.1cm}
\plotone{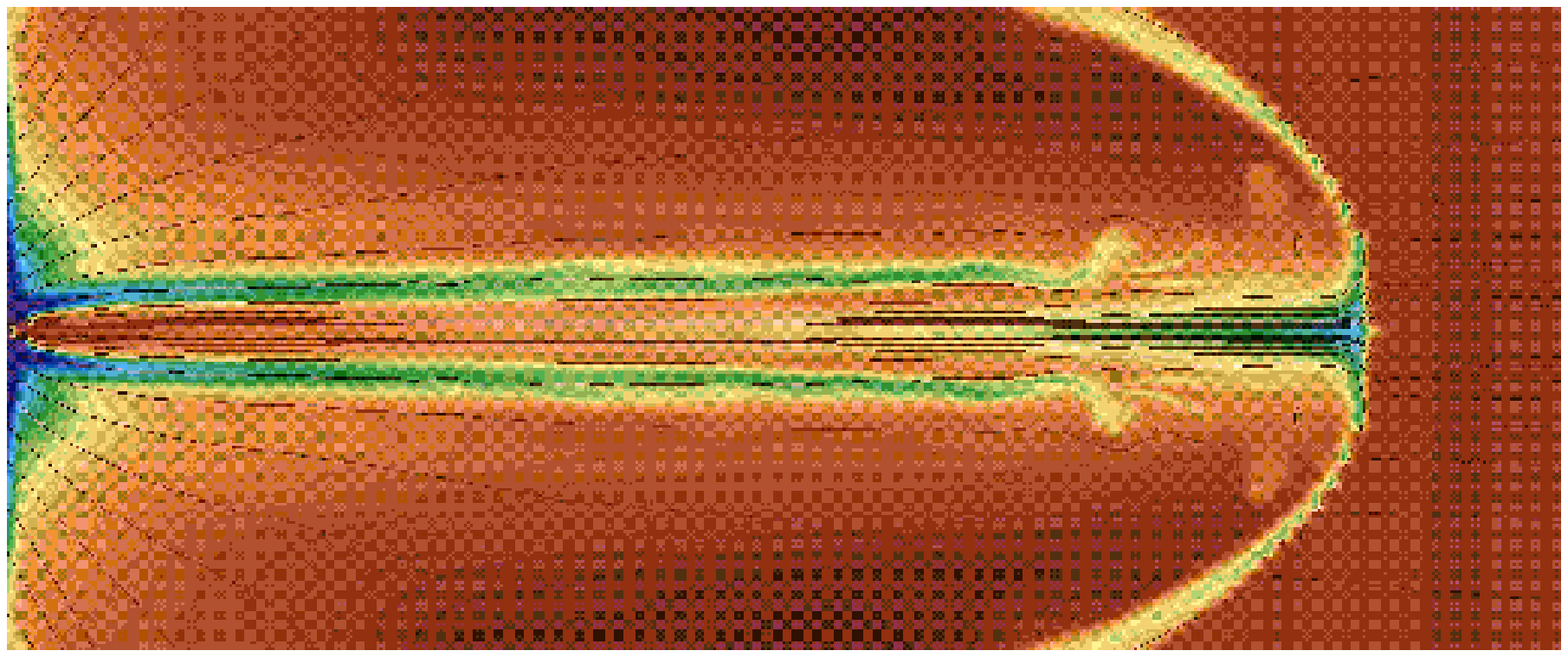}\\
\vspace{.1cm}
\plotone{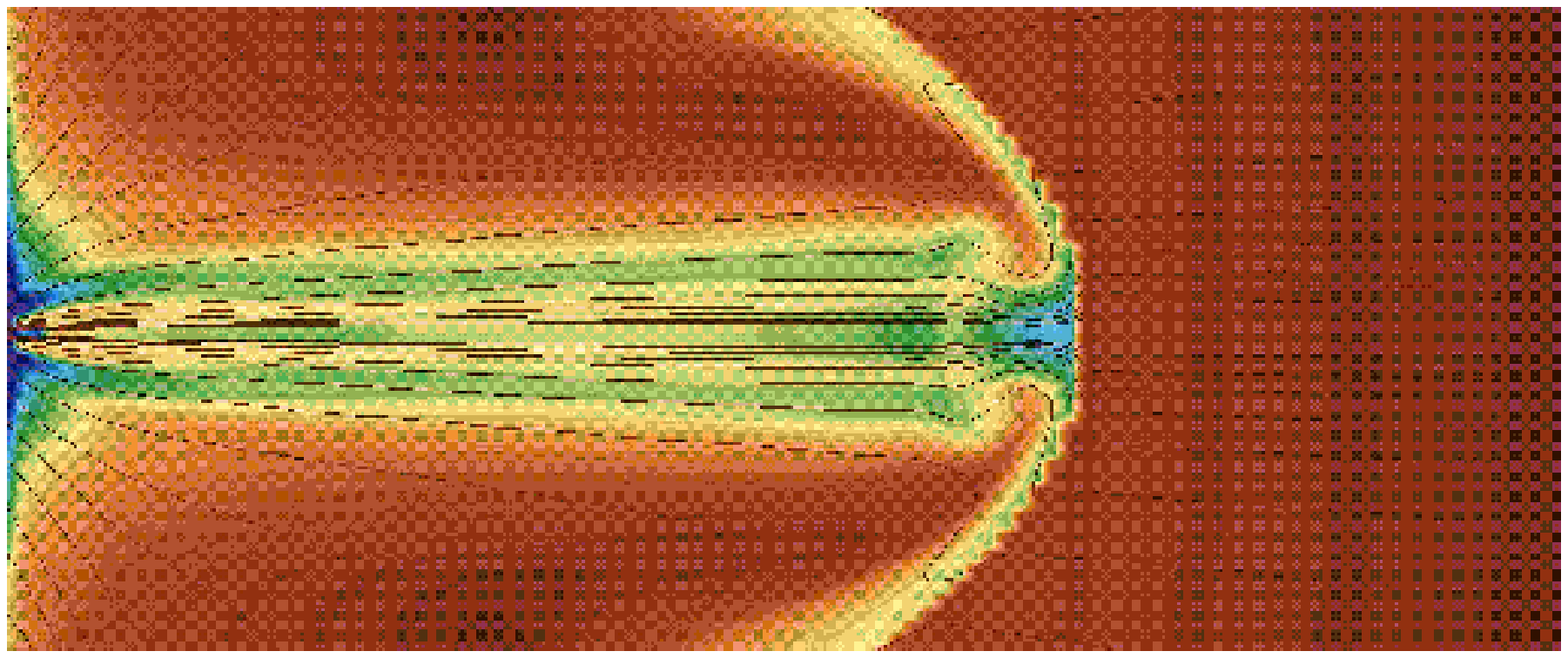}

\caption{Global evolution of the jet on a grid of 
(Z$\times$R)=(280$\times$40)R$_{\rm i}$ with a resolution 
of 900$\times$200 elements for different magnetic diffusivity. $\eta$=0,0.01,0.1 for top, middle and bottom panel, respectively. Colors (blue to yellow in decreasing manner) indicate density and lines denote linearly spaced poloidal field lines. The bow shock advances slower with increasing magnetic diffusivity.}
\end{figure}

We perform two types of simulations: {\bf 1)} The disk is defined as time-independent boundary condition (Ouyed \& Pudritz 1997; Fendt \& Elstner 2000; Fendt \& \v{C}emelji\'{c} 2002). Here we investigate the acceleration and collimation of a resistive jet. {\bf 2)} The evolution of the disk structure is included in the numerical simulation.

\section{Formation of resistive jets}

Within a subgrid of 280$\times$80 grid points and physical grid of  (Z$\times$R)=(140$\times$40)R$_{\rm i}$ we investigated the transition from  accretion of matter in the resistive disk to ejection in the jet for up to 4000 rotations of the inner disk radius R$_{\rm i}$. The magnetic diffusivity, in combination with the magnetic and inertial forces and pressure and gravity, modifies the MHD structure of the jet. In our simulations the jet velocity inccreases with increasing magnetic diffusivity. The same holds for the degree of collimation. We defined the mass flux across the surfaces parallel to the accretion disk boundaries:
\[\dot{M_z}=\int^{r_{\rm max}}_0 2\pi r\rho u_zdr\ ,\ \dot{M_r}=\int^{z_{\rm max}}_0 2\pi r\rho u_rdz\ .\]
As a measure of the collimation we propose the ratio of the radial to the axial mass flow, where for this ratio $>$1 the flow becomes decollimated. We found that below the 'critical' $\eta$  (in our choice of parameters $\eta<0.5$), the outflow is collimated. 

\begin{figure}
{\plotfiddle{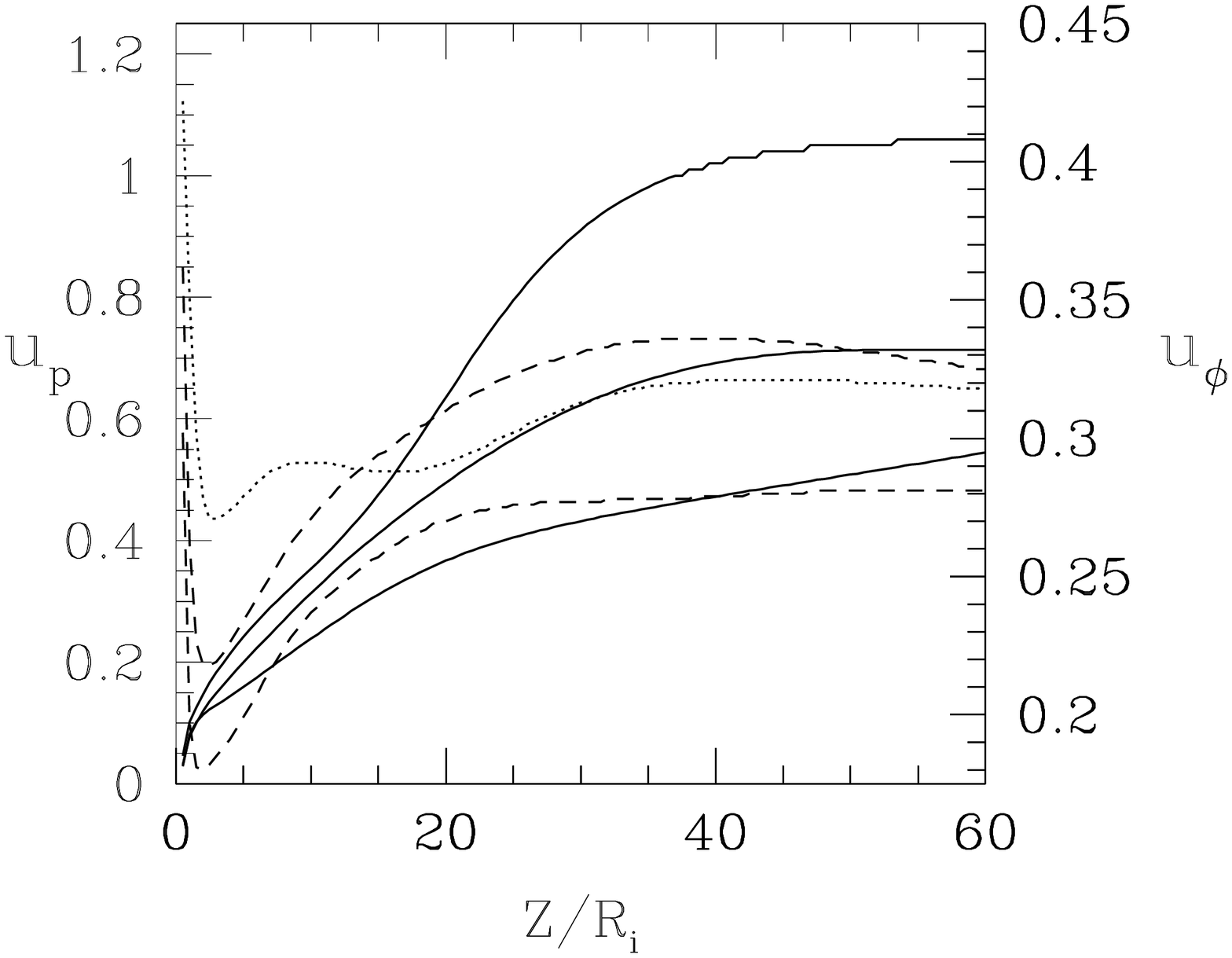}{4cm}{0}{31}{31}{0}{-110}}{\plotfiddle{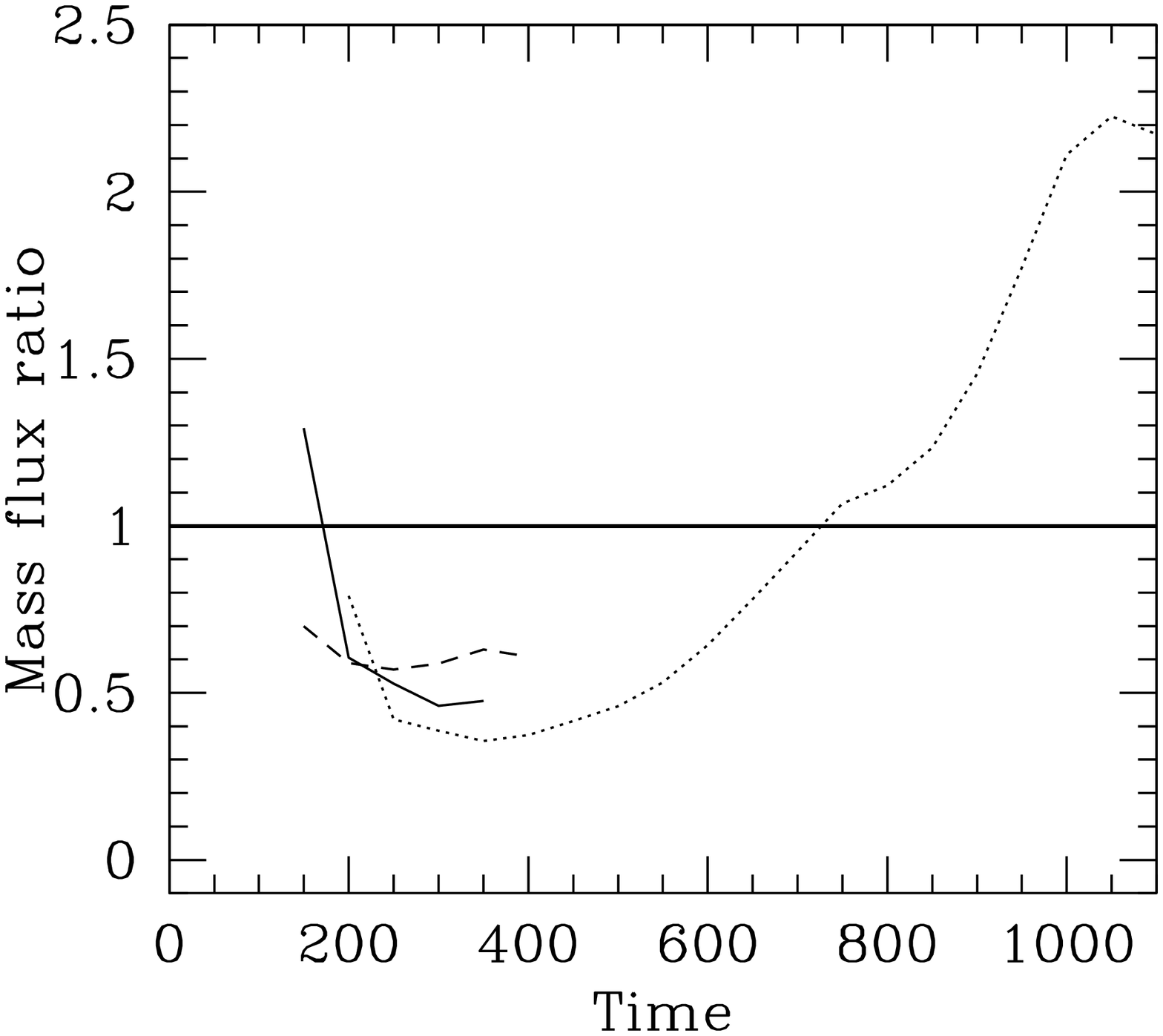}{0cm}{0}{28}{28}{200}{-60}}
\vspace{.5cm}
\caption{Components of the velocities and time evolution of the mass flow ratio. Left panel: Cuts in Z direction at R=15 for the quasistationary time of simulations. Poloidal components are given in solid lines, toroidal in dashed lines (note the different left and right labels). Right panel:  The radial/axial mass flow in the inner part of the jet, (Z$\times$R)=(60$\times$20)R$_{\rm i}$. Parameter $\eta$=0,0.1,0.5 in solid, dashed and dotted line, respectively. The degree of collimation decreases with increasing $\eta$.
}
\end{figure}

 The toroidal field may lead to (de-)accelerating Lorentz forces $\vec{F}_{\rm L, ||} \sim \vec{j}_{\perp} \times \vec{B}_{\phi} $ and (de-)collimating forces
$\vec{F}_{L,\perp} \sim \vec{j}_{||} \times \vec{B}$ (projections with respect to the poloidal magnetic field), see Figure 3.

\begin{figure}
\plottwo{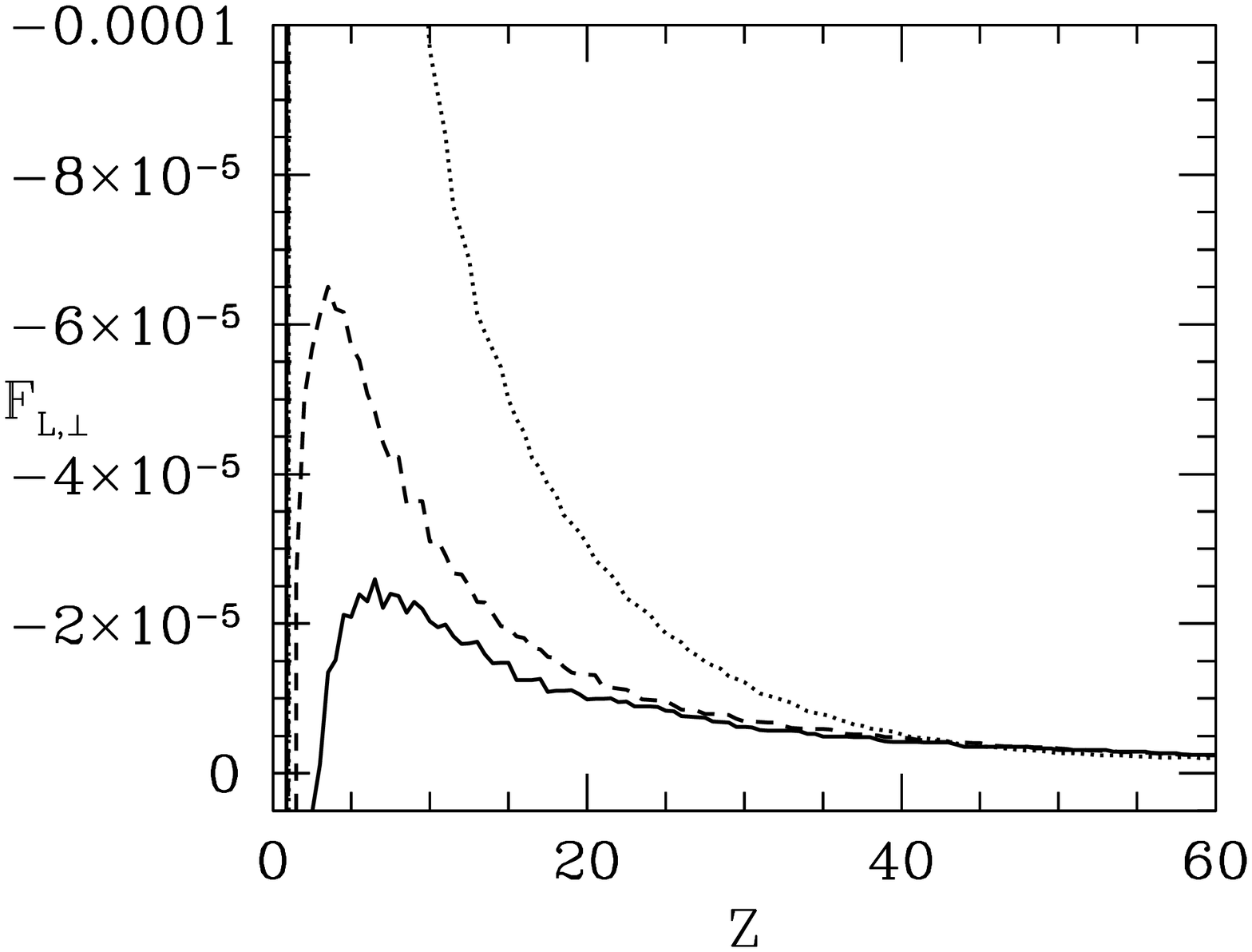}{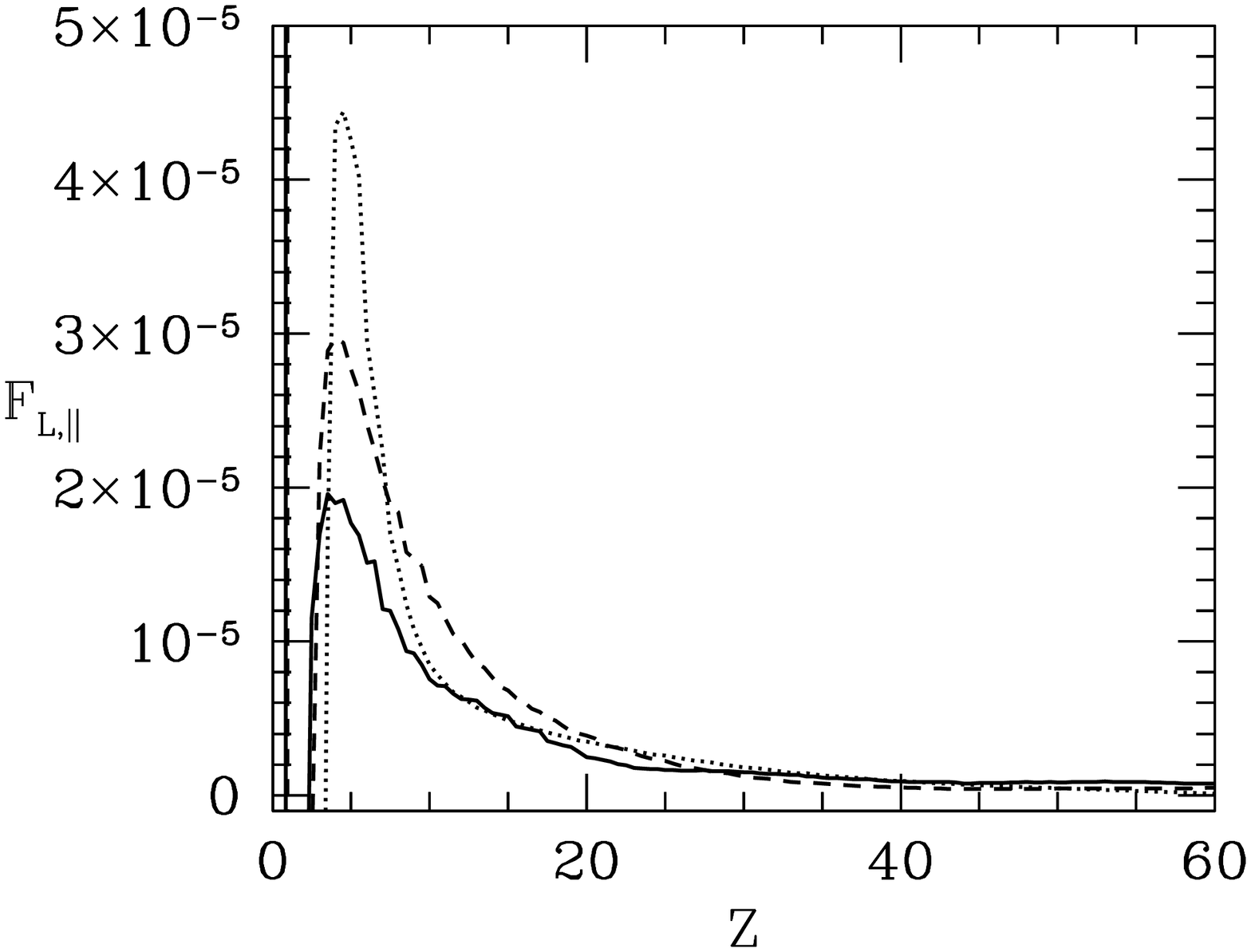}
\vspace{-.5cm}
\caption{Lorentz forces in the jet for different magnetic diffusivity $\eta$ = 0, 0.1, 0.5 (solid, dashed and dotted lines, respectively) along a flux surface leaving the box close to the (R=20,Z=60)-corner. For the perpendicular component the positive sign denotes the R-direction (de-collimating force). For the parallel component the positive sign denotes the Z-direction (accelerating force).
}
\end{figure}

As a de-collimation of the poloidal magnetic structure also implies a smaller launching angle for the sub-Alfv\'enic flow, the magneto-centrifugal acceleration mechanism may work more efficiently.

\section{Jet \& disk evolution}

The vertical disk equilibrium and the Keplerian rotation imposes the
constraint for the radial density profile. The initial poloidal velocity
profile is defined only by the accretion rate in the disk. The magnetic
field is vertically uniform and the magnetic diffusivity is of a Gaussian
profile. All the other boundaries are defined as open.

\begin{figure}
\plottwo{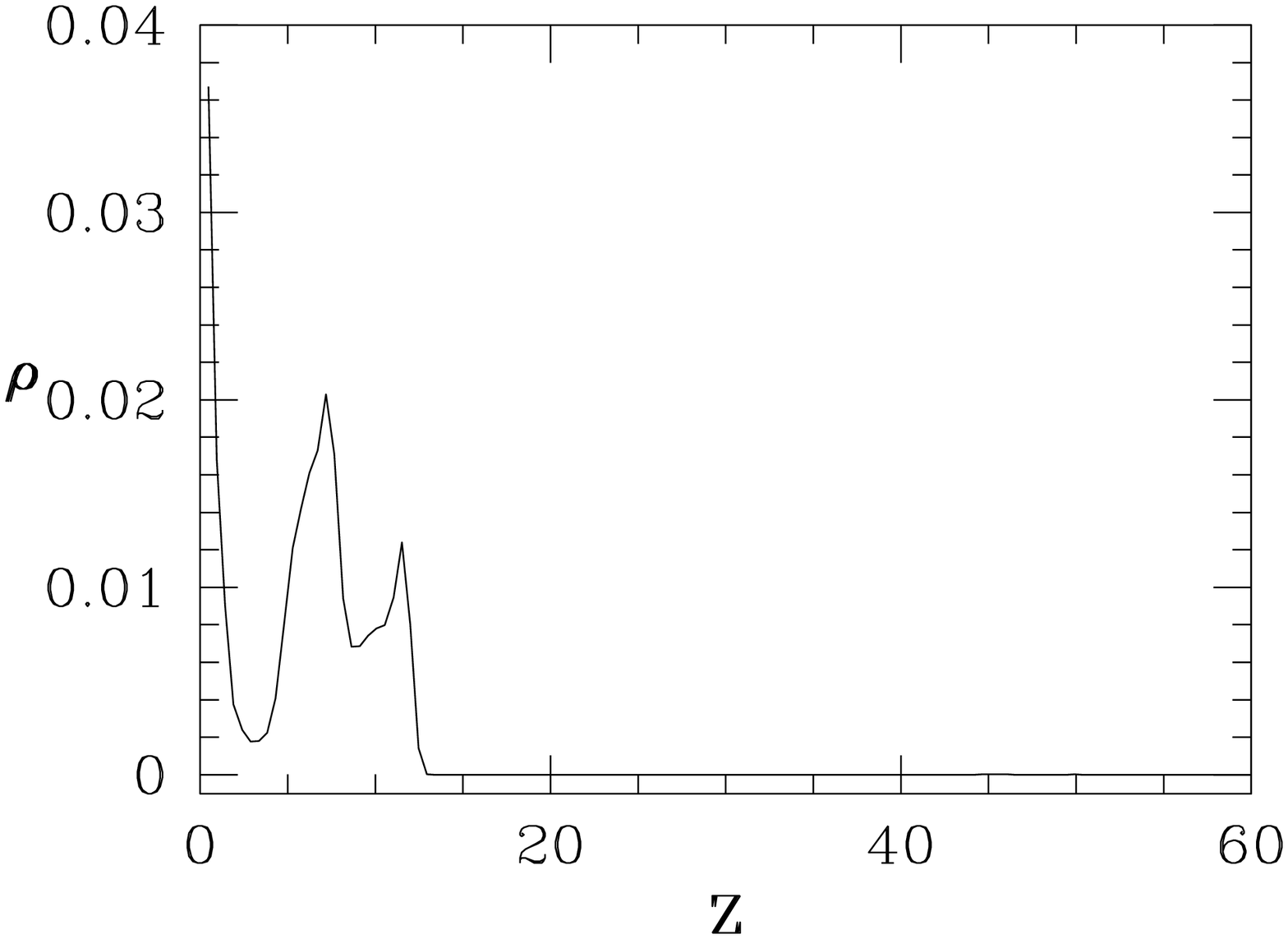}{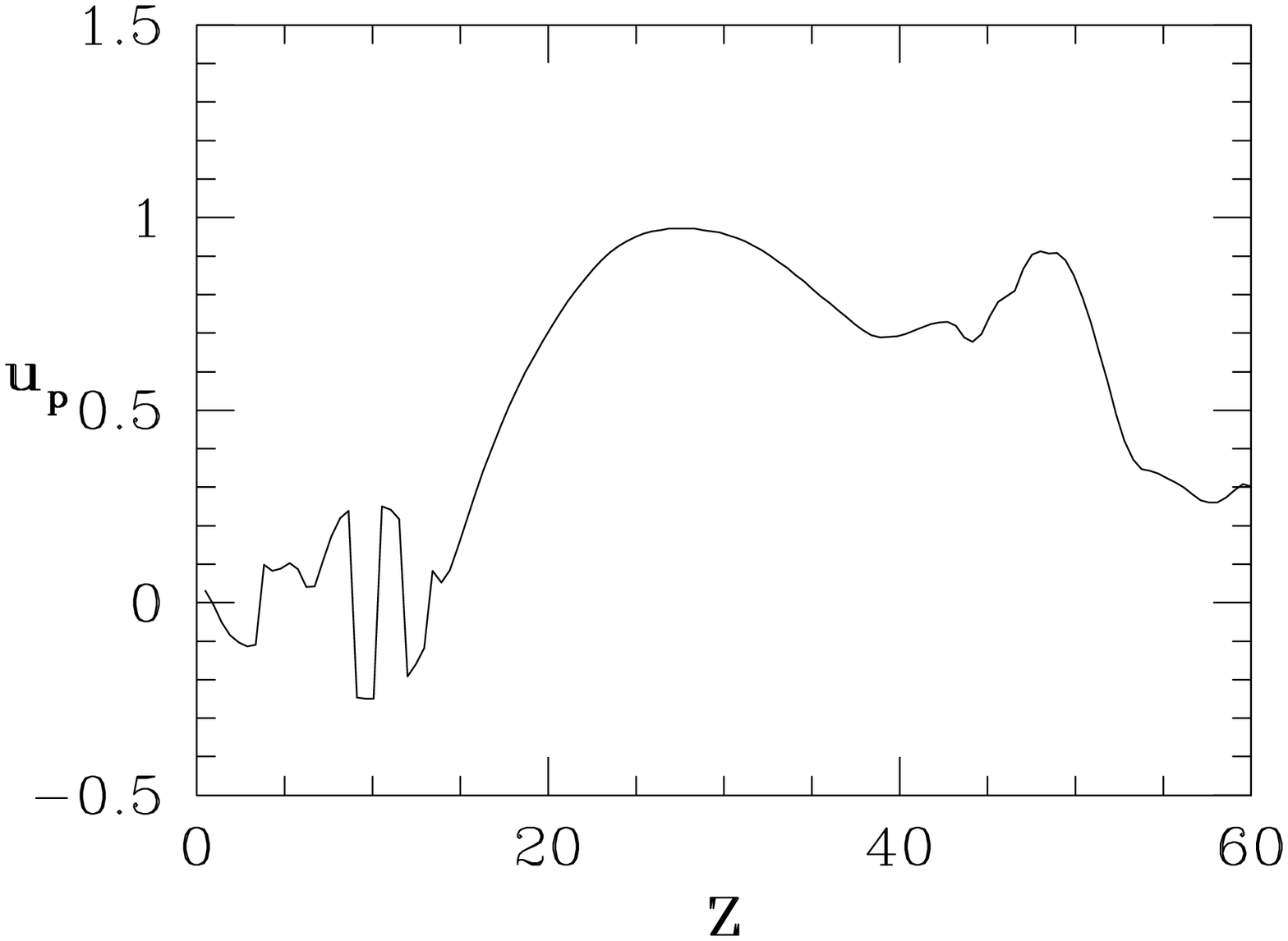}
\vspace{-1.2cm}
\caption{The density and the poloidal velocity. Cut in Z direction at R=15.}
\end{figure}

Close to the origin, a 'sink' region is defined, ensuring the outflow to
contain only the matter accreted from the disk, and not re-bounced from the
symmetry axis of the jet (Casse \& Keppens 2002).

\begin{figure}
{\plotfiddle{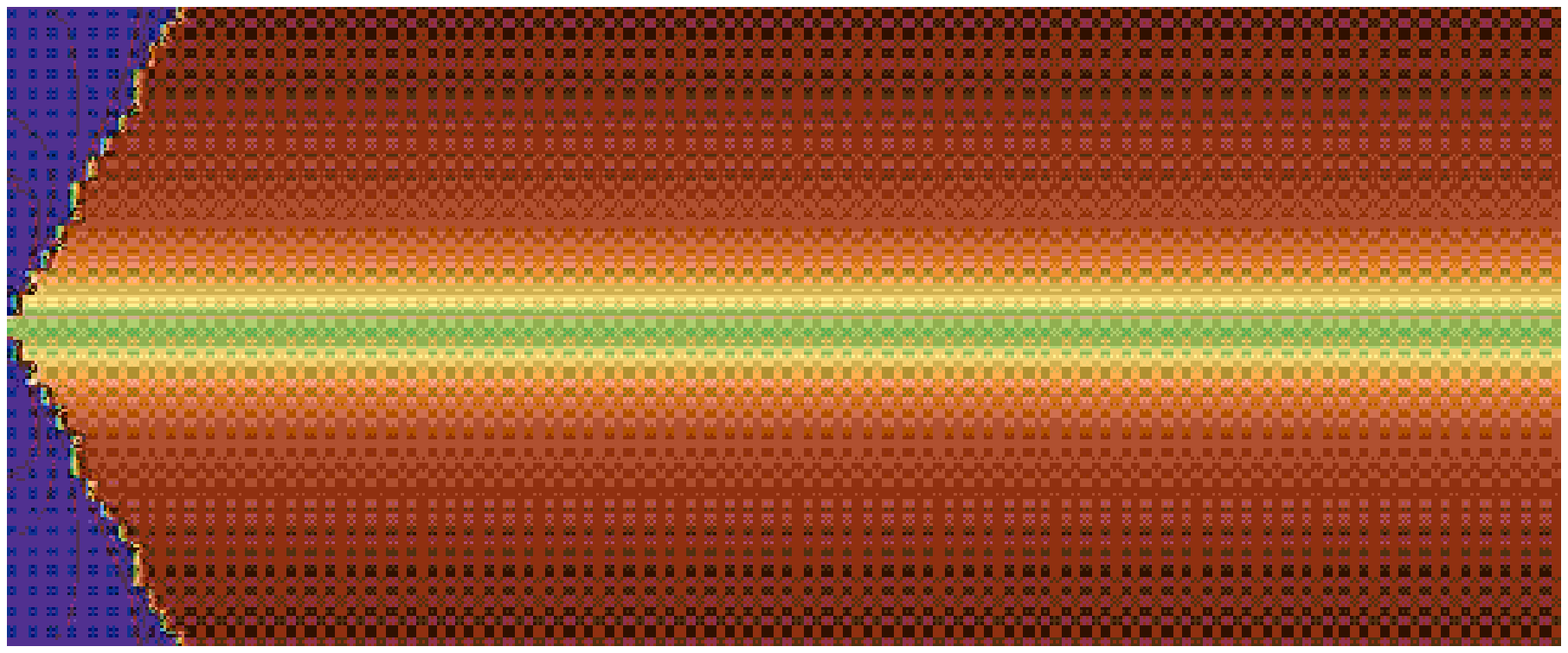}{2cm}{0}{35}{50}{0}{-38}}{\plotfiddle{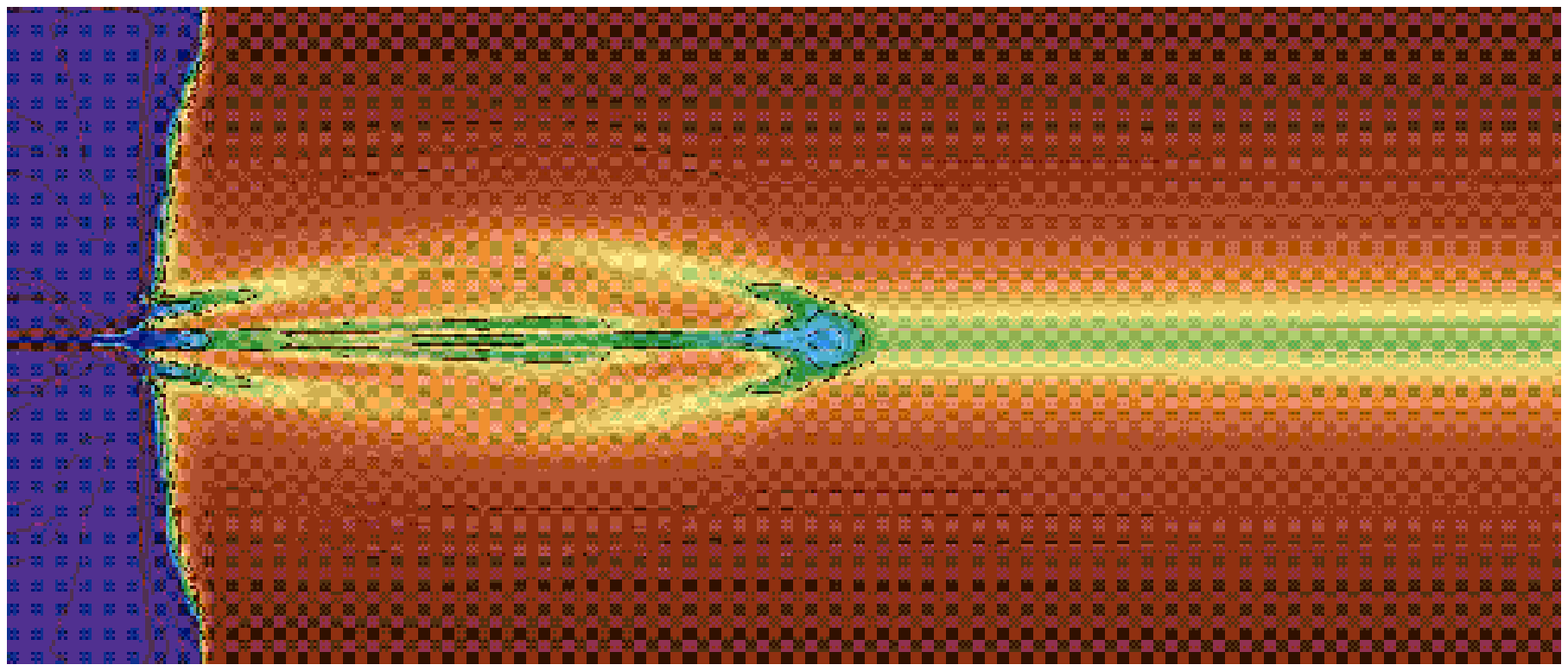}{0cm}{0}{35}{50}{190}{-27}}\\
{\plotfiddle{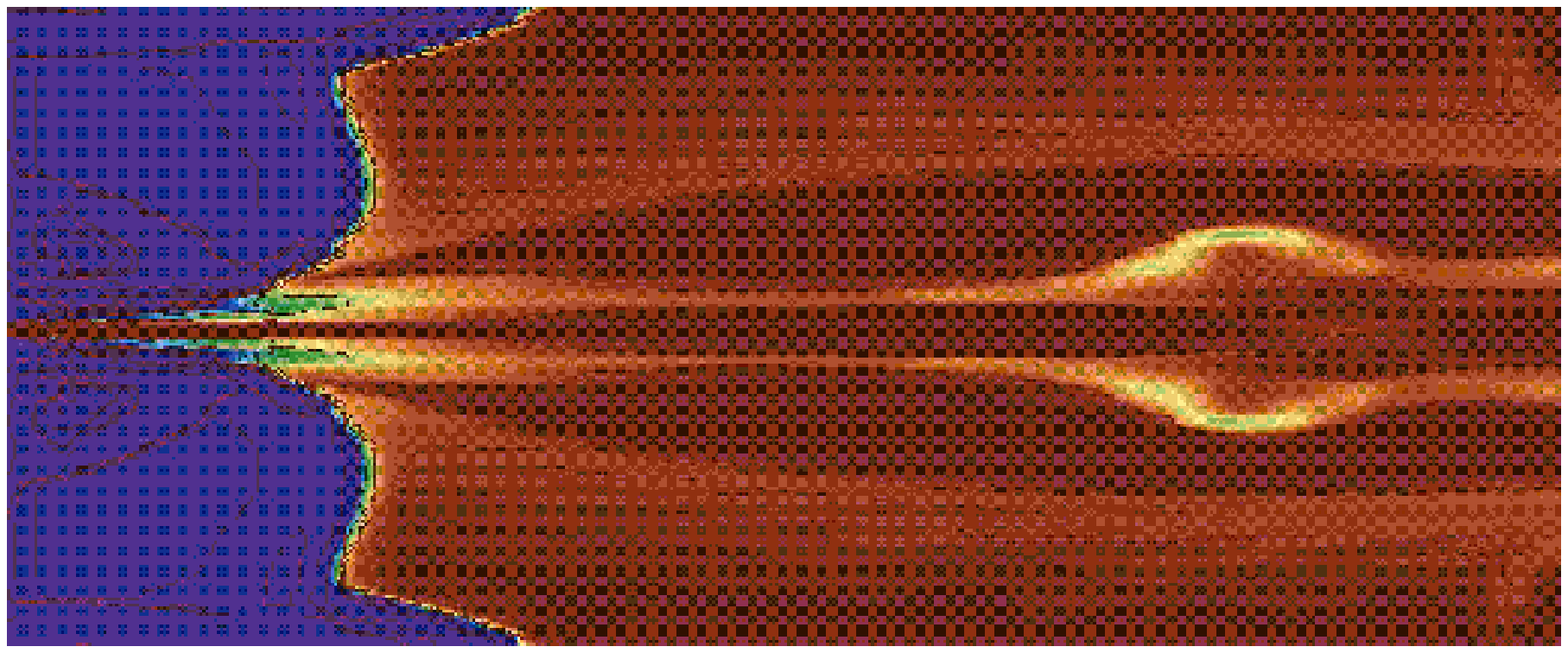}{2cm}{0}{35}{50}{0}{-78}}

\vspace{2.7cm}
\caption{Evolution of the outflow from the accretion disk on a grid of (Z$\times$R)=(60$\times$40)R$_{\rm i}$ in a mesh of 125$\times$80
 elements for t=0,50,150 rotations of the disk inner radii. Colors (blue to yellow in decreasing manner) indicate
 density and lines are linearly spaced poloidal field lines.
}
\end{figure}

Our preliminary results are as follows (Figure 5). The jet launching occurs
only within R$<$5, indicating the existence of the conditions for the
ejection only in the inner part of the jet (plasma $\beta$ and angle of the
magnetic field lines decrease for increasing radii), in agreement with Casse
\& Keppens (2002). The outflow is well collimated, with the velocity equal
to the escape velocity of the central object. There exists strong indication
for stationarity of the outflow after few tens of rotations. The jet radius
is R$\sim$20. The disk piles-up for a factor two to three in comparison to
the initial height, reaching new equilibrium.

\begin{acknowledgments}
This work was partly financed by the  DFG Schwerpunktprogramm ``Physik der Sternentstehung''(FE490/2-1). M.\v{C}. thanks the IAU for the travel grant for this meeting.
\end{acknowledgments}

\end{document}